\documentclass{jpsj2}
\usepackage{graphicx}

\title{Three-Dimensional Multiband d-p Model of Superconductivity in Spin-Chain Ladder Cuprate}

\author{Shigeru Koikegami
\footnote{E-mail address : shigeami@secondlab.co.jp} 
and Takashi Yanagisawa$^1$}

\inst{Second Lab, LLC, 
2-32-1 Umezono, Tsukuba, Ibaraki 305-0045\\
$^1$Nanoelectronics Research Institute, 
AIST Tsukuba Central 2, Tsukuba, Ibaraki 305-8568}

\abst{We study the superconductivity in the three-dimensional multiband d-p model, 
in which a Cu$_2$O$_3$-ladder layer and a CuO$_2$-chain layer are alternately stacked, 
as a model of the superconducting spin-chain ladder cuprate. $p_z$-Wave-like triplet superconductivity 
is found to be the most stable, and its dependence on interlayer coupling can explain the superconducting 
transition temperature dependence on pressure in real superconducting spin-chain ladder cuprates. 
The superconductivity may be enhanced if hole transfer from the chain layer to the ladder layer can 
be promoted beyond the typical transfer rate.}

\kword{superconductivity, three-dimensional multiband d-p model, spin-chain ladder cuprate, 
interlayer coupling}

\begin{document}
\sloppy
\maketitle

\section{Introduction}
The spin-chain ladder cuprate Sr$_{14-x}$Ca$_x$Cu$_{24}$O$_{41}$ (Sr14-24-41) has attracted 
much interest as a related material of high-$T_{\mathrm{c}}$ cuprates for more than two decades. 
It has three types of layers, known as the chain, spacer, and ladder.\cite{McCarron1988,Siegrist1988,Kato1994} 
These layers are stacked in an alternating manner along the layer axis $b$. 
The chain layer contains CuO$_2$ chains, while the ladder layer consists of two-leg Cu$_2$O$_3$ ladders. 
Theoretically, the two-leg spin-$1/2$ antiferromagnetic Heisenberg ladder is a system 
with a finite energy gap for reaching the lowest magnetic excitation.\cite{Dagotto1996} 
The hole-doped two-leg spin ladder has been investigated theoretically 
and confirmed numerically to possess an instability towards $d$-wave-like superconductivity 
on the basis of the $t$-$J$ ladder model,
\cite{Dagotto1992,Rice1993,Sigrist1994,Tsunetsugu1994,Hayward1995,Jeckelmann1998} one-band Hubbard ladder model,
\cite{Yamaji1994,Khveshchenko1994,Noack1994,Asai1994,Yanagisawa1995,Balents1996,Kuroki1996,Dahm1997,Jeckelmann1998,Koike1999} 
and three-band Hubbard ladder model.\cite{Jeckelmann1998,Nishimoto2002} 
In 1996, Sr$_{14-x}$Ca$_x$Cu$_{24}$O$_{41}$ with $x=13.6$ was found to be superconducting 
under high pressures, $P \geq 3$GPa.\cite{Uehara1996} In Sr14-24-41, 
doped holes move from the chain layer to the ladder layer by Ca substitution for Sr.\cite{Kato1996} 
High pressures intensify this self-doping effect and afford the electrons in the ladder layer 
sufficient itinerancy.\cite{Motoyama1997,Osafune1997} However, the superconducting transition 
temperature ($T_{\mathrm{c}}$) has a maximum at $P \approx 5$GPa and then decreases with increasing 
pressure.\cite{Isobe1998} 
The dependence of $T_{\mathrm{c}}$ on pressure is common to other compositions as well, such as 
Sr$_{14-x}$Ca$_x$Cu$_{24}$O$_{41}$ for $x=11.5$~\cite{Nagata1998} 
and $x=10,11,12$.\cite{Eisaki2000,Motoyama2002} 
Such superconductivity behavior is similar to that of high-$T_{\mathrm{c}}$ cuprates in the overdoped regime. 
Considering the experimental results of the electronic properties, 
the superconductivity of Sr14-24-41 can be thought of as an extension of that of 
high-$T_{\mathrm{c}}$ cuprates.\cite{Kojima2001}

In accord with the above novel results, 
the two-leg Hubbard ladders coupled via a weak interladder hopping were investigated theoretically 
by Kishine and Yonemitsu.\cite{Kishine1997} According to their perturbative renormalization group analysis, 
the system has a $d$-wave-like superconducting ground state, and restores the interladder 
coherence within an increase in the extent of interladder hopping. 
Moreover, the superconducting state and its $T_{\mathrm{c}}$ were 
evaluated on the basis of the two-dimensional (2D) Trellis-lattice Hubbard model, 
i.e., the coupled two-leg ladder model.\cite{Kontani1998,Sasaki2004,Kuroki2005} 
In their studies, a wide doping regime exists in which $d$-wave like singlet superconductivity appears. 
In one of them,\cite{Sasaki2004} it is also shown that a certain doping regime prefers $p_z$-wave like triplet superconductivity. 
Thus, it is plausible that the superconductivity may be intrinsically linked to the 2D two-leg ladder structure.
Practically, however, the superconductivity in Sr14-24-41 appears for certain compounds 
only under high pressures. In order to understand this situation, 
we should find routes for enhancing superconductivity in real compounds, that have not been 
considered yet in past theoretical studies. It is worth noting that 
Isobe et al. have already pointed out the effects of the hybrid orbital 
between Cu $3d$ in the ladder layer and O $2p$ in the chain layer on the superconductivity 
under high pressures.\cite{Isobe1998,Isobe2000} 
According to their arguments, the hybrid orbital accelerates 
the redistribution of holes and enhances the superconductivity.

In this study, we investigate the superconductivity of the two-leg ladder layer coupled 
with the chain layer. We adopt the three-dimensional (3D) d-p model with the quasi-one-dimensional (Q1D) 
structure in which a Cu$_2$O$_3$ ladder layer and a CuO$_2$ chain layer are alternately
stacked. In our model, the ladder and chain layers are coupled via 
hybridization between the Cu $3d$ orbital in the ladder layer and the O $2p$ orbital in the chain layer. 
Moreover, we introduce a sufficiently small on-site Coulomb interaction showing 
that the second-order perturbation theory (SOPT) 
can be justified. We can treat the superconductivity using a 
weak coupling analysis because, in our model, the effective interaction for 
Cooper pairing is so small that only electrons on the Fermi surface 
are involved in the superconductivity. In particular, the weak coupling formulation 
by Kondo is applicable even in the case with a very small effective interaction.\cite{JKondo2001} 
We previously applied Kondo's formulation to the study of the 
3D d-p model with multilayer perovskite structure 
and investigated how the superconducting gap depends on the number of layers.\cite{Koikegami2006} 
As in the study of multilayer cuprates, we show that calculation on the basis of a 3D model 
is practical for assessing the superconductivity in spin-chain ladder cuprate. 
Our results give a possible explanation as to why the superconductivity appears in 
Sr$_{14-x}$Ca$_x$Cu$_{24}$O$_{41}$ for $x \geq 10$ only under high pressures.

\section{Formulation}
\begin{figure}
\includegraphics[width=8.0cm]{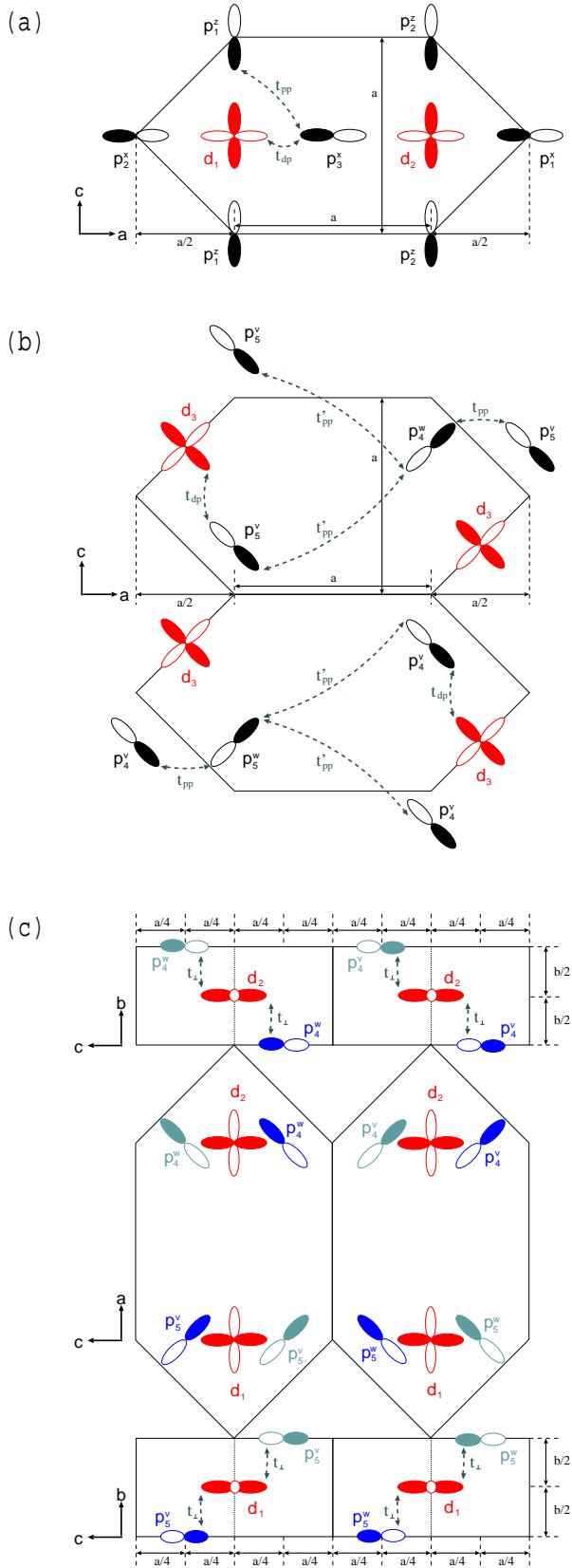}
\caption{\label{figure:1}(Color online) Transfer energies in the 3D chain-ladder d-p model: (a) within the ladder layer, 
(b) within the chain layer, and (c) for the interlayer hopping.}
\end{figure}
Our 3D d-p model is schematically shown in Fig.~\ref{figure:1}. 
We consider two Cu sites and three O sites in the ladder layer, and one Cu site and two O sites in the chain layer. 
In the chain layer, the lattice constants along the c-axis and a-axis are $\sqrt{2}$ and $1/\sqrt{2}$ 
times those of Sr14-24-41, respectively. Owing to this technique, we do not need a large unit cell, 
which is needed in crystal analysis of Sr14-24-41, in order to treat the interlayer hopping. 
We can decompose our 3D d-p model composed of the ladder layer and chain layer into several parts as follows:
\begin{align}
H  = & H_{\mathrm{Ladder}}^0+H_{\mathrm{Chain}}^0+H_{\mathrm{Ladder-Chain}}^0
+H^\prime \nonumber \\
& -\mu \sum_{\mib{k} \sigma} \left[d_{1\mib{k}\sigma}^\dagger d_{1\mib{k}\sigma}+d_{2\mib{k}\sigma}^\dagger d_{2\mib{k}\sigma}
+p_{1\mib{k}\sigma}^{z \dagger}p_{1\mib{k}\sigma}^z+p_{1\mib{k}\sigma}^{x \dagger}p_{1\mib{k}\sigma}^x
+p_{2\mib{k}\sigma}^{z \dagger}p_{2\mib{k}\sigma}^z+p_{2\mib{k}\sigma}^{x \dagger}p_{2\mib{k}\sigma}^x
+p_{3\mib{k}\sigma}^{x \dagger}p_{3\mib{k}\sigma}^x \right.\nonumber \\
& \hspace{4em}\left.+d_{3\mib{k}\sigma}^\dagger d_{3\mib{k}\sigma}
+p_{4\mib{k}\sigma}^{w \dagger}p_{4\mib{k}\sigma}^w+p_{4\mib{k}\sigma}^{v \dagger}p_{4\mib{k}\sigma}^v
+p_{5\mib{k}\sigma}^{w \dagger}p_{5\mib{k}\sigma}^w+p_{5\mib{k}\sigma}^{v \dagger}p_{5\mib{k}\sigma}^v\right],
\label{eq:11}
\end{align}
where $d_{l\mib{k}\sigma}$ ($d_{l\mib{k}\sigma}^{\dagger}$)  
and $p_{m\mib{k}\sigma}^\nu$ ($p_{m\mib{k}\sigma}^{\nu\dagger}$)  are 
the annihilation (creation) operators for d-electrons in the $l$-th site and for 
p$^\nu$-electrons in the $m$-th site, 
having a momentum $\mib{k}$ and spin $\sigma=\{\uparrow,\downarrow\}$, respectively. 
The site indices $l$ and $m$, 
and the orbital index $\nu$ are defined as shown in Fig.~\ref{figure:1}. 
In the following, we take both $a$ and $b$ as the 
unit of length and put $a=b=1$. $\mu$ represents the chemical potential. 
The noninteracting parts in eq.~(\ref{eq:11}), 
i.e., $H_{\mathrm{Ladder}}^0$,  $H_{\mathrm{Chain}}^0$, and $H_{\mathrm{Ladder-Chain}}^0$, are represented by
\begin{align}
\hspace*{-26pt}
H_{\mathrm{Ladder}}^0 = & \nonumber \\
 & \hspace{-5em}
\sum_{\mib{k} \sigma}\left(d_{1\mib{k}\sigma}^\dagger\,d_{2\mib{k}\sigma}^\dagger\,
p_{1\mib{k}\sigma}^{z \dagger}\,p_{1\mib{k}\sigma}^{x \dagger}\,
p_{2\mib{k}\sigma}^{z \dagger}\,p_{2\mib{k}\sigma}^{x \dagger}\,
p_{3\mib{k}\sigma}^{x \dagger}\right)\!
\left(
\begin{array}{ccccccc} 
\varepsilon^{dp} & 0 & \zeta_{\mib{k}}^{z*} & 0 & 0 & \zeta_{\mib{k}}^{x*} & -\zeta_{\mib{k}}^x \\
0 & \varepsilon^{dp} & 0 & -\zeta_{\mib{k}}^x & \zeta_{\mib{k}}^{z*} & 0 & \zeta_{\mib{k}}^{x*} \\ 
\zeta_{\mib{k}}^z & 0 & 0 & 0 & 0 & \zeta_{\mib{k}}^{p*} & \zeta_{\mib{k}}^p \\
0 & -\zeta_{\mib{k}}^{x*} & 0 & 0 & \zeta_{\mib{k}}^{p*} & 0 & 0 \\
0 & \zeta_{\mib{k}}^z & 0 & \zeta_{\mib{k}}^p & 0 & 0 & \zeta_{\mib{k}}^{p*} \\
\zeta_{\mib{k}}^x & 0 & \zeta_{\mib{k}}^p & 0 & 0 & 0 & 0 \\
-\zeta_{\mib{k}}^{x*} & \zeta_{\mib{k}}^x & \zeta_{\mib{k}}^{p*} & 0 & \zeta_{\mib{k}}^p & 0 & 0 
\end{array} \right)\!
\left(\begin{array}{c} 
d_{1\mib{k}\sigma} \\
d_{2\mib{k}\sigma} \\
p_{1\mib{k}\sigma}^z\\
p_{1\mib{k}\sigma}^x\\
p_{2\mib{k}\sigma}^z\\
p_{2\mib{k}\sigma}^x\\
p_{3\mib{k}\sigma}^x
\end{array} \right)\!,
\nonumber \\
\label{eq:2}
\end{align}
\begin{align}
\hspace*{-29pt}
H_{\mathrm{Chain}}^0 = 
\sum_{\mib{k} \sigma}\left(d_{3\mib{k}\sigma}^\dagger\,
p_{4\mib{k}\sigma}^{w \dagger}\,p_{4\mib{k}\sigma}^{v \dagger}\,
p_{5\mib{k}\sigma}^{w \dagger}\,p_{5\mib{k}\sigma}^{v \dagger}\right)\!\left(
\begin{array}{ccccc} 
\varepsilon^{dp}-\Delta V & -\xi_{\mib{k}}^w & \xi_{\mib{k}}^{v*} & \xi_{\mib{k}}^{w*} & -\xi_{\mib{k}}^v \\  
-\xi_{\mib{k}}^{w*} & -\Delta V & \xi_{\mib{k}}^{z*} & 0 & \xi_{\mib{k}}^{x*} \\ 
\xi_{\mib{k}}^v & \xi_{\mib{k}}^z & -\Delta V & \xi_{\mib{k}}^{x*} & 0 \\
\xi_{\mib{k}}^w & 0 & \xi_{\mib{k}}^x & -\Delta V & \xi_{\mib{k}}^z \\
-\xi_{\mib{k}}^{v*} & \xi_{\mib{k}}^x & 0 & \xi_{\mib{k}}^{z*} & -\Delta V 
\end{array} \right)\!
\left(\begin{array}{c} 
d_{3\mib{k}\sigma} \\
p_{4\mib{k}\sigma}^w\\
p_{4\mib{k}\sigma}^v\\
p_{5\mib{k}\sigma}^w\\
p_{5\mib{k}\sigma}^v
\end{array} \right)\!,
\nonumber \\
\label{eq:8}
\end{align}
and
\begin{equation}
H_{\mathrm{Ladder-Chain}}^0 = \sum_{\mib{k} \sigma}\left[\eta_{\mib{k}}^+ 
\left(p_{4\mib{k}\sigma}^{v \dagger} d_{2\mib{k}\sigma}-p_{4\mib{k}\sigma}^{w \dagger} d_{2\mib{k}\sigma}\right)
+\eta_{\mib{k}}^- \left(p_{5\mib{k}\sigma}^{w \dagger} d_{1\mib{k}\sigma}-p_{5\mib{k}\sigma}^{v \dagger} d_{1\mib{k}\sigma}\right)
+{\mathrm{H.c.}}\right],
\label{eq:9}
\end{equation}
respectively. In eqs.~(\ref{eq:2})--(\ref{eq:9}), we use the abbreviations 
$\zeta_{\mib{k}}^z=2{\mathrm{i}}t_{dp}\sin\frac{k_z}{2}$, 
$\zeta_{\mib{k}}^x=t_{dp}e^{-{\mathrm{i}}k_x/2}$, 
$\zeta_{\mib{k}}^p=2{\mathrm{i}}t_{pp}e^{-{\mathrm{i}}k_x/2}\sin\frac{k_z}{2}$, 
$\xi_{\mib{k}}^w=t_{dp}e^{{\mathrm{i}}k_z/2}e^{{\mathrm{i}}k_x/4}$, 
$\xi_{\mib{k}}^v=t_{dp}e^{{\mathrm{i}}k_z/2}e^{-{\mathrm{i}}k_x/4}$, 
$\xi_{\mib{k}}^z=t_{pp}e^{{\mathrm{i}}k_z}$, 
$\xi_{\mib{k}}^x=-t_{pp}e^{{\mathrm{i}}k_x/2}-2t_{pp}^\prime e^{-{\mathrm{i}}k_x}\cos\frac{k_z}{2}$, and 
$\eta_{\mib{k}}^{\pm}=2\mathrm{i}t_\perp \sin\left(\frac{k_y}{2}\pm\frac{k_z}{4}\right)$. The transfer energies, 
$t_{dp}$, $t_{pp}$, $t_{pp}^\prime$, and $t_\perp$ are defined as shown in Fig.~\ref{figure:1}. The increase in 
$t_\perp$ in our model is considered to represent the increase in pressure in real Sr14-24-41. 
$\varepsilon^{dp}$ is the level difference between d- and p-electrons. Moreover, we use $\Delta V$ 
to control the charge imbalance between the ladder and chain layers. 
The change in $\Delta V$ can represent the change in Madelung energy due to the Ca doping of Sr14-24-41. 
Considering only the on-site Coulomb repulsion among d-electrons, 
the interacting part $H^\prime$ in eq.~(\ref{eq:11}) is described as
\begin{equation}
H^\prime = \frac{U}{N} \sum_{l=1}^3 \sum_{\mib{k} \mib{k}^\prime \mib{q}}
        d_{l\mib{k}+\mib{q}\uparrow}^{\dagger} 
	d_{l\mib{k}^\prime-\mib{q}\downarrow}^{\dagger} 
	d_{l\mib{k}^\prime\downarrow} 
	d_{l\mib{k}\uparrow}.
\label{eq:10}
\end{equation}
In eq.~(\ref{eq:10}), $N$ is the number of $\mib{k}$-space lattice points in the first Brillouin zone (FBZ).

In the following analysis, we assume that only electrons on the Fermi surface 
of the same band can have pair instability. For our 3D d-p model, 
$2$ or $3$ $d$-like bands intersect with the Fermi level. Thus, according to 
the Bardeen-Cooper-Schrieffer (BCS) theory, we have the following self-consistent equation for the pair 
function on the $\lambda$-th $d$-like band, $\Phi_{\mib{k}}^\lambda$:
\begin{equation}
\Phi_{\mib{k}}^\lambda=
-\frac{1}{2N}\!\sum_{ij\nu\mib{k}^\prime}%
V_{ij}(\mib{k}+\mib{k}^\prime)
\frac{z_i^\lambda(\mib{k})z_j^\nu(\mib{k}^\prime)}
{\sqrt{\left(\varepsilon_{\mib{k}^\prime}^\nu-\mu \right)^2+
\left(\Phi_{\mib{k}^\prime}^\nu\right)^2}}\,\Phi_{\mib{k}^\prime}^\nu,
\label{eq:3}
\end{equation}
where $i,j=1,2,3$ (Cu site indices) and $\lambda,\nu=1,2(,3)$ 
($d$-like band indices). This equation is valid for both the spin-singlet and spin-triplet pair functions 
if we define $V_{ij}(\mib{q})$ differently as the need arises. 
Thus, hereafter, we omit the spin indices. $V_{ij}(\mib{q})$ represents the effective 
pair scattering between a d-electron on the $i$-th site and one 
on the $j$-th site. The term $\varepsilon_{\mib{k}}^\nu$ represents the energy 
dispersion of the $\nu$-th $d$-like band, and $z_i^\lambda(\mib{k})$ 
represents the matrix element of unitary transformation. 
These variables are obtained by solving the eigen-equation 
for the noninteracting part $H_{\mathrm{Ladder}}^0+H_{\mathrm{Chain}}^0+H_{\mathrm{Ladder-Chain}}^0$ in eq.~(\ref{eq:11}).
We set $\Phi_{\mib{k}}^\lambda=\Delta_{\mathrm{sc}}\!\cdot\!\Psi_{\mib{k}}^\lambda$, 
where $\Delta_{\mathrm{sc}}$ denotes the magnitude of $\Phi_{\mib{k}}^\lambda$, 
and $\Psi_{\mib{k}}^\lambda$ represents its $\mib{k}$ dependence on the $\lambda$-th $d$-like band. 
On the basis of Kondo's argument,\cite{JKondo2001} retaining only the divergent term, we can rewrite eq.~(\ref{eq:3}) as 
\begin{equation}
\Psi_{\mib{k}}^\lambda = \log_e \Delta_{\mathrm{sc}}\cdot \frac{1}{N}
\sum_{ij\nu \mib{k}^\prime}V_{ij}(\mib{k}+\mib{k}^\prime)
 z_i^\lambda(\mib{k})z_j^\nu(\mib{k}^\prime)
\delta(\varepsilon_{\mib{k}^\prime}^\nu-\mu)
\Psi_{\mib{k}^\prime}^\nu,
\label{eq:4}
\end{equation}
for a very small $\Delta_{\mathrm{sc}}$. Equation (\ref{eq:4}) 
is a homogeneous integral equation for 
$\Psi_{\mib{k}}^\lambda$ with an eigen-value of $1/\log \Delta_{\mathrm{sc}}$. 
We are interested in obtaining the most stable pairing state, 
so we must find the eigenvector $\Psi_{\mib{k}}^\lambda$ with the smallest
eigenvalue $1/\log \Delta_{\mathrm{sc}}$ using eq.~(\ref{eq:4}) 
when $\Delta_{\mathrm{sc}}$ is maximum. Given the quasi-two-dimensionality of 
our 3D d-p model, i.e., $t_{\perp} \ll t_{dp}$, we assume five functions as candidates for 
the most stable pairing state: 
\begin{align}
& \hspace{2.5em}a^\lambda : \mbox{$s$-wave-like singlet}, \\ 
& \sum_{m_z=1}^M a_z^\lambda(m_z)\cos m_zk_z+\sum_{m_z=1}^M \sum_{m_x=1}^{2m_z}a_x^\lambda(m_zm_x)\cos \frac{2m_z-m_x}{2}k_z\cos \frac{3m_x}{2}k_x+\sum_{m_y=1}^M a_y^\lambda(m_y)\cos m_yk_y \nonumber \\
& \hspace{3.5em}: \mbox{$d_{z^2-x^2}$-wave-like singlet},  \\ 
& \sum_{m_z=1}^M \sum_{m_x=1}^{2m_z-1}a^\lambda(m_zm_x)\sin \frac{2m_z-m_x}{2}k_z\sin \frac{3m_x}{2}k_x : \mbox{$d_{zx}$-wave-like singlet}, \nonumber \\ 
& \sum_{m_z=1}^M a^\lambda(m_z)\sin m_zk_z : \mbox{$p_z$-wave-like triplet}, \label{eq:13}\\
& \sum_{m_z=1}^M \sum_{m_x=1}^{2m_z}a^\lambda(m_zm_x)\cos \frac{2m_z-m_x}{2}k_z\sin \frac{3m_x}{2}k_x : \mbox{$p_x$-wave-like triplet}. 
\end{align}
In order to solve eq.~(\ref{eq:4}), we substitute these candidates for 
$\Psi_{\mib{k}}^\lambda$ and $\Psi_{\mib{k}^\prime}^\nu$, 
and integrate for $k_z$, $k_z^\prime$, $k_x$, $k_x^\prime$, $k_y$, and $k_y^\prime$. 
Then, we can safely reduce our original eigenvalue problem 
for $\Psi_{\mib{k}}^\lambda$ to an eigenvalue problem for 
$a^\lambda$ [$a_z^\lambda(m_z)$, $a_x^\lambda(m_zm_x)$, and $a_y^\lambda(m_y)$] in order to obtain the most stable pairing state. 
When we solve it numerically by the standard method, we obtain both the eigenvalue 
$1/\log_e \Delta_{\mathrm{sc}}$ and the eigenvector $a^\lambda$ [$a_z^\lambda(m_z)$, $a_x^\lambda(m_zm_x)$, and $a_y^\lambda(m_y)$]. 
Here, to solve eq.~(\ref{eq:4}) within SOPT for singlet-pairing states, we have 
\begin{equation}
V_{ij}(\mib{q}) = U\delta_{ij}+\frac{U^2}{N}\sum_{\kappa\rho\mib{k}}
z_i^\kappa(\mib{q}+\mib{k})z_j^\rho(\mib{k})
\frac{\left(1-f_{\mib{q}+\mib{k}}^\kappa\right)f_{\mib{k}}^\rho}
{\varepsilon_{\mib{q}+\mib{k}}^\kappa-\varepsilon_{\mib{k}}^\rho}.
\label{eq:1}
\end{equation}
Meanwhile, for triplet-pairing states,
\begin{equation}
V_{ij}(\mib{q}) = -\frac{U^2}{N}\sum_{\kappa\rho\mib{k}}
z_i^\kappa(\mib{q}+\mib{k})z_j^\rho(\mib{k})
\frac{\left(1-f_{\mib{q}+\mib{k}}^\kappa\right)f_{\mib{k}}^\rho}
{\varepsilon_{\mib{q}+\mib{k}}^\kappa-\varepsilon_{\mib{k}}^\rho},
\label{eq:12}
\end{equation}
where
\begin{equation}
f_{\mib{k}}^\rho = \frac{1}{2}
\left[1-\tanh\left(\frac{\varepsilon_{\mib{k}}^\rho-\mu}{2T}
\right)\right],
\label{eq:7}
\end{equation} 
and $T$ denotes the temperature.

\section{Results and Discussion}
In our present analyses, all $\varepsilon_{\mib{k}}^\nu$ and $z_i^\lambda(\mib{k})$ in eq.~(\ref{eq:4}) 
are first calculated for the $\mib{k}-$points on an equally spaced mesh in FBZ for each band. 
The mesh size along $k_z$ is $108$, and the sizes along $k_x$ and $k_y$ are both $64$. 
Then, we calculate $V_{ij}(\mib{k}+\mib{k}^\prime)$ in eq.~(\ref{eq:4}) only for $\mib{k}-$ and 
$\mib{k}^\prime-$points satisfying the conditions 
$\varepsilon_{\mib{k}}^\lambda=\mu$ and $\varepsilon_{\mib{k}^\prime}^\nu=\mu$, respectively. 
When we calculate $V_{ij}(\mib{k}+\mib{k}^\prime)$ according to eqs.~(\ref{eq:1}) and (\ref{eq:7}), 
we set the temperature $T=0.001\,$eV$\approx 10\,$K, at which our system can be considered to behave similarly 
to the ground state. These calculations have been performed at $U=0.3\,$eV, 
where magnetic instabilities cannot occur. Other common parameters are $t_{dp}=1.00\,$eV, 
$t_{pp}=-0.50\,$eV, $t_{pp}^\prime=-0.10\,$eV, and $\varepsilon^{dp}=2.60\,$eV. 
These parameters are determined using 
examples from studies of the three-band Hubbard ladder model,\cite{Jeckelmann1998,Nishimoto2002} 
local-density approximation,\cite{Arai1997} and angle-resolved photo-emission spectroscopy (ARPES).
\cite{Takahashi1997,Sato1998,TYoshida2009}

As mentioned in the last section, varying the pressure and Ca doping in real Sr14-24-41 can be reproduced 
by changing $t_\perp$ and $\Delta V$, respectively, in our model. Thus, in order to comprehensively understand 
the superconductivity of spin-chain ladder cuprate, we calculate the $\log_e \Delta_{\mathrm{sc}}$ of 
the most stable superconducting state for various $t_\perp$ and $\Delta V$ values. 
All through these calculations, 
we fix average holes for Cu$^{2+}$ $[n^{\mathrm{h}}(\mathrm{Cu}^{2+})]$ to $0.25$ as well as the value of 
real Sr14-24-41.\cite{McCarron1988} 
Here, $n^{\mathrm{h}}(\mathrm{Cu}^{2+})$ is defined as 
$n^{\mathrm{h}}(\mathrm{Cu}^{2+}) \equiv [2n^{\mathrm{h}}_{\mathrm{Ladder}}(\mathrm{Cu}^{2+})+n^{\mathrm{h}}_{\mathrm{Chain}}(\mathrm{Cu}^{2+})]/3$, 
where $n^{\mathrm{h}}_{\mathrm{Ladder}}(\mathrm{Cu}^{2+})$ and $n^{\mathrm{h}}_{\mathrm{Chain}}(\mathrm{Cu}^{2+})$ are average holes per Cu$^{2+}$ 
in the ladder and chain layers, respectively. These values should be estimated as the total holes in our model in order to be 
compared with those of real Sr14-24-41.\cite{McCarron1988} We find that the most stable superconducting state of our five candidates 
is always the $p_z$-wave-like triplet, represented by eq.~(\ref{eq:13}). We confirm that $\log_e \Delta_{\mathrm{sc}}$ estimated 
for this state with $M=16$ does not differ from that with $M=15$ by more than $8\%$.  
The following discussion is therefore restricted to the $p_z$-wave-like triplet state with $M=16$.

\begin{figure}
\includegraphics[width=8.0cm]{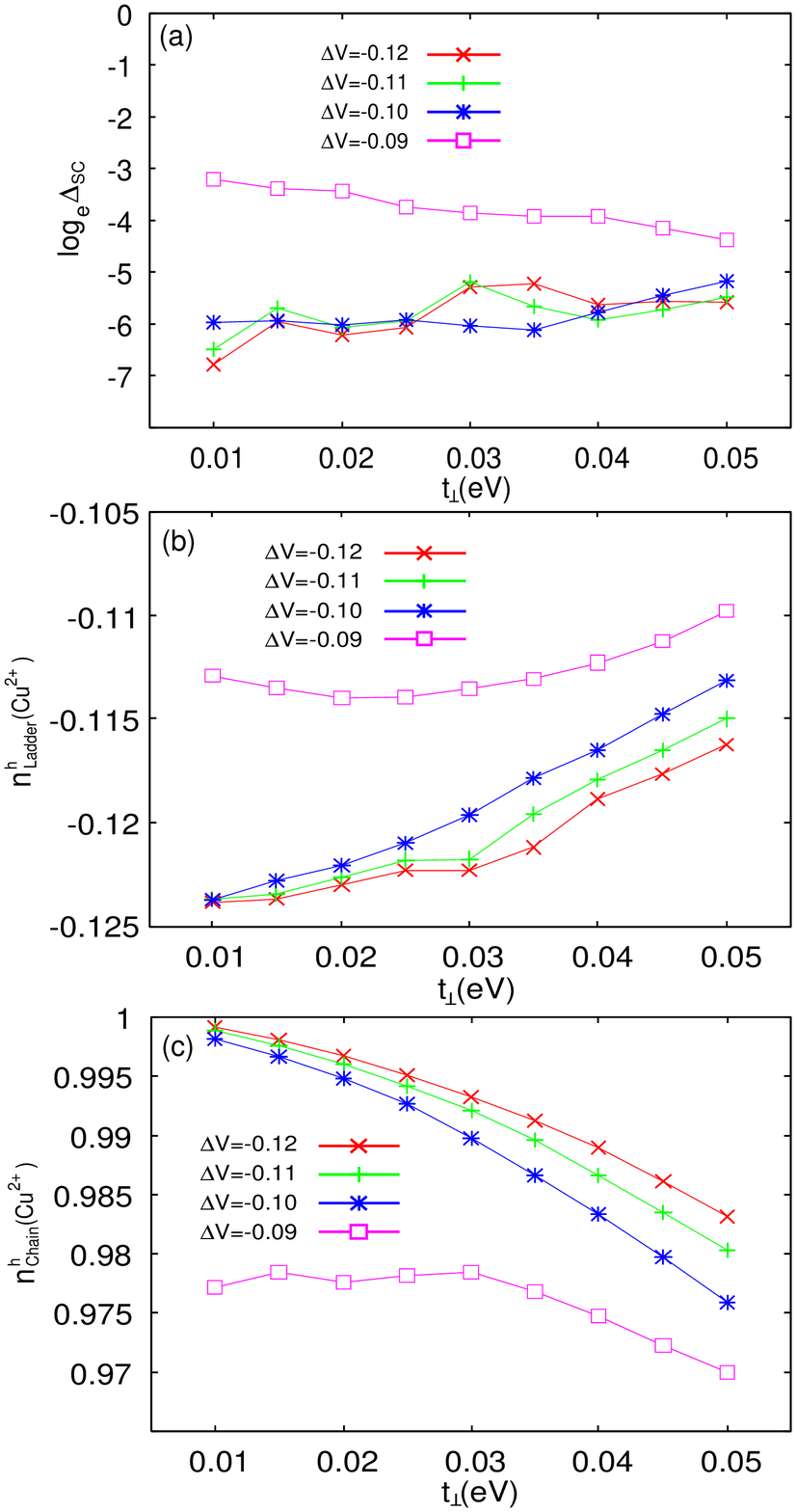}
\caption{\label{figure:2}(Color online) (a) $\log \Delta_{\mathrm{sc}}$, 
(b) $n^{\mathrm{h}}_{\mathrm{Ladder}}(\mathrm{Cu}^{2+})$, and (c) $n^{\mathrm{h}}_{\mathrm{Chain}}(\mathrm{Cu}^{2+})$ 
for $\Delta V=-0.12$, $-0.11$, $-0.10$, and $-0.09\,$eV.}
\end{figure}

In Fig.~\ref{figure:2}, we summarize how $\log_e \Delta_{\mathrm{sc}}$, 
$n^{\mathrm{h}}_{\mathrm{Ladder}}(\mathrm{Cu}^{2+})$ and $n^{\mathrm{h}}_{\mathrm{Chain}}(\mathrm{Cu}^{2+})$ 
depend on $t_\perp$ for various $\Delta V$ values. 
Figure~\ref{figure:2}(a) shows the $\log_e \Delta_{\mathrm{sc}}$ dependence on $t_\perp$ for 
$\Delta V=-0.12\,$eV and $\Delta V=-0.11\,$eV, as well as the $T_{\mathrm{c}}$ dependence on pressure in real 
Sr14-24-41,\cite{Isobe1998,Nagata1998,Eisaki2000,Motoyama2002} 
i.e., it has a maximum at an intermediate $t_\perp$. Since the temperature $T=0.001\,$eV$\approx 10\,$K 
is thought to be sufficiently low, $T_{\mathrm{c}}$ should be proportional to $\Delta_{\mathrm{sc}}(T)$ within the BCS theory. 
Thus, our results of the $\log_e \Delta_{\mathrm{sc}}$ dependence on $t_\perp$ for 
$\Delta V=-0.12\,$eV and $\Delta V=-0.11\,$eV can qualitatively 
reproduce the $T_{\mathrm{c}}$ dependence on pressure in real Sr14-24-41. 
Moreover, as shown in Figs.~\ref{figure:2}(b) and \ref{figure:2}(c), Cu$^{2+}$ holes are transferred 
from the chain layer to the ladder layer with an increase in $t_\perp$, 
which also qualitatively agrees with the experimental results under high pressures.\cite{Isobe1998,Isobe2000} 

\begin{figure}
\includegraphics[width=8.0cm]{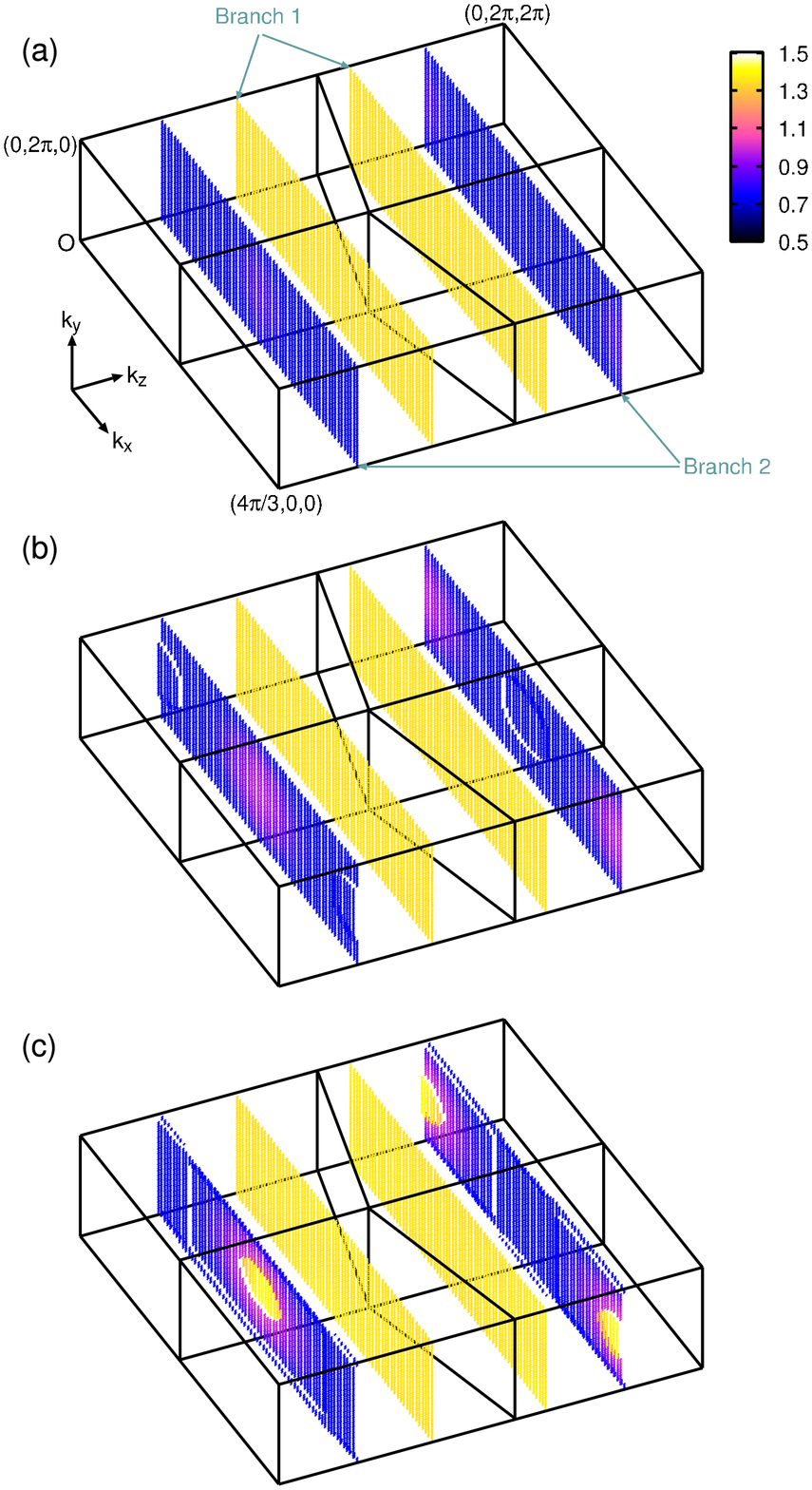}
\caption{\label{figure:3}(Color online) Densities of states (DOS's) on Fermi surface for $\Delta V=-0.12\,$eV: (a) for $t_\perp=0.010\,$eV, 
(b) for $t_\perp=0.030\,$eV, and (c) for $t_\perp=0.050\,$eV.}
\end{figure}

\begin{figure}
\includegraphics[width=8.0cm]{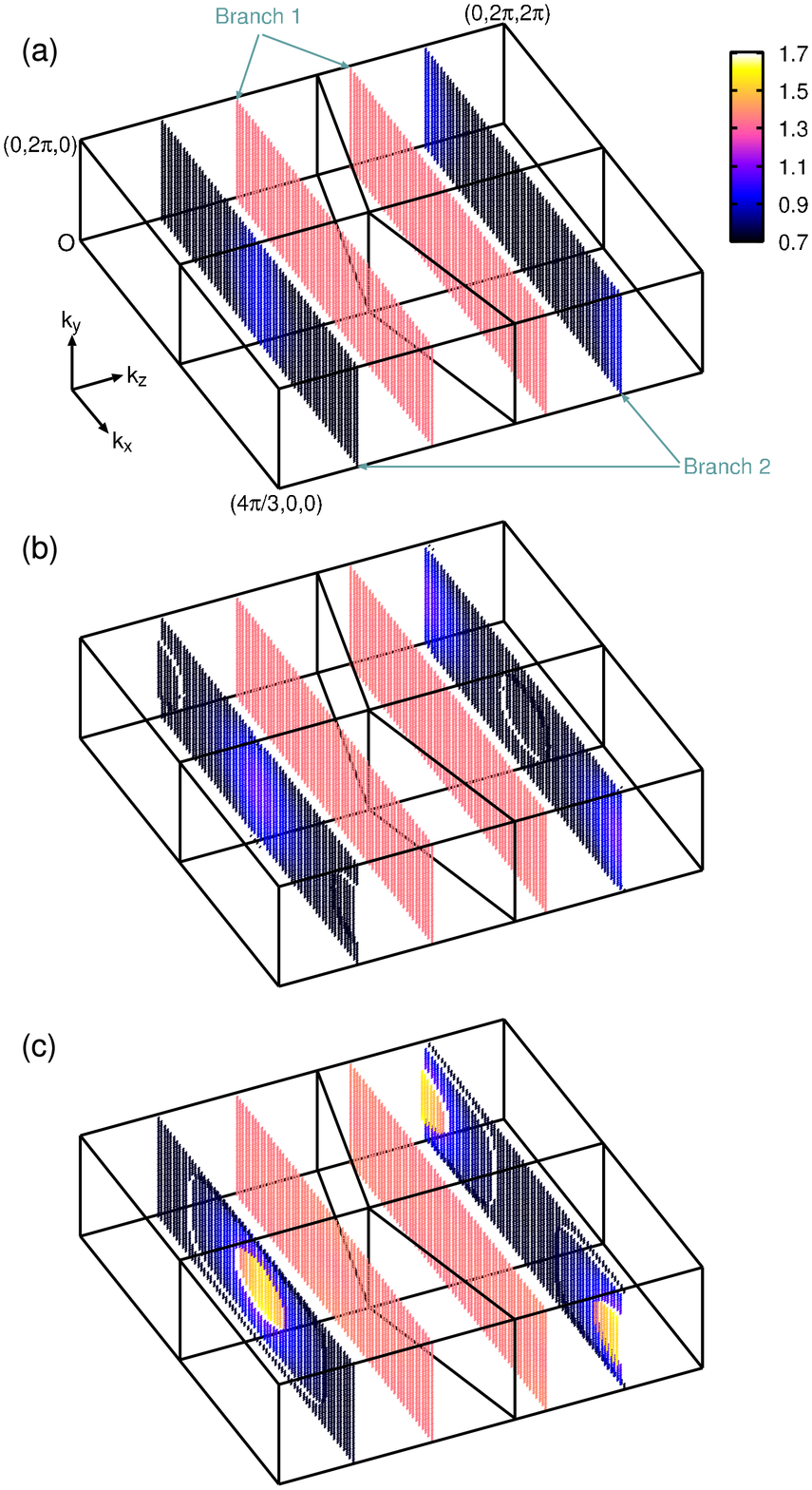}
\caption{\label{figure:4}(Color online) DOS's on Fermi surface for $\Delta V=-0.11\,$eV: (a) for $t_\perp=0.010\,$eV, 
(b) for $t_\perp=0.030\,$eV, and (c) for $t_\perp=0.050\,$eV.}
\end{figure}

\begin{figure}
\includegraphics[width=8.0cm]{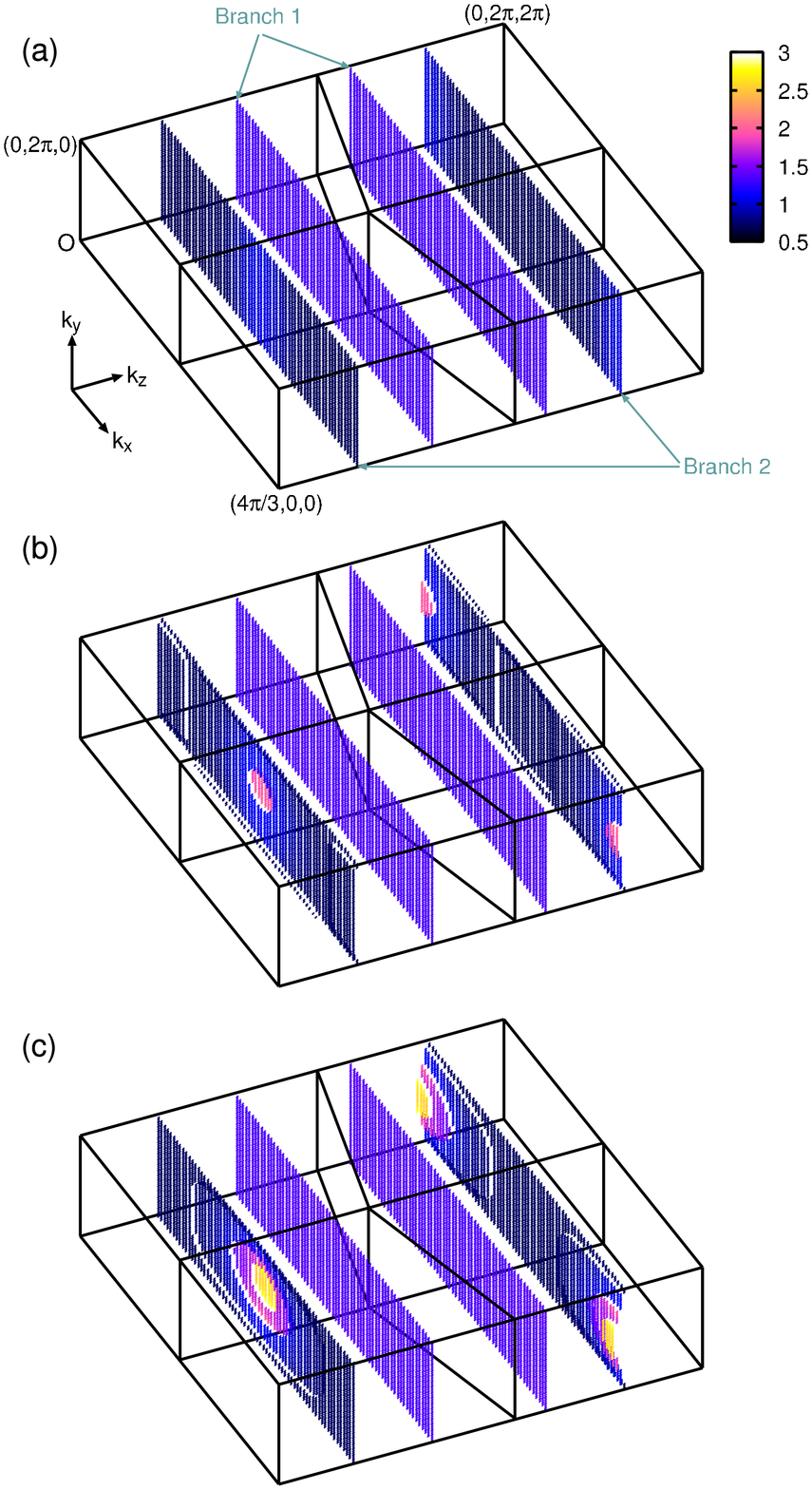}
\caption{\label{figure:5}(Color online) DOS's on Fermi surface for $\Delta V=-0.10\,$eV: (a) for $t_\perp=0.010\,$eV, 
(b) for $t_\perp=0.030\,$eV, and (c) for $t_\perp=0.050\,$eV.}
\end{figure}

\begin{figure}
\includegraphics[width=8.0cm]{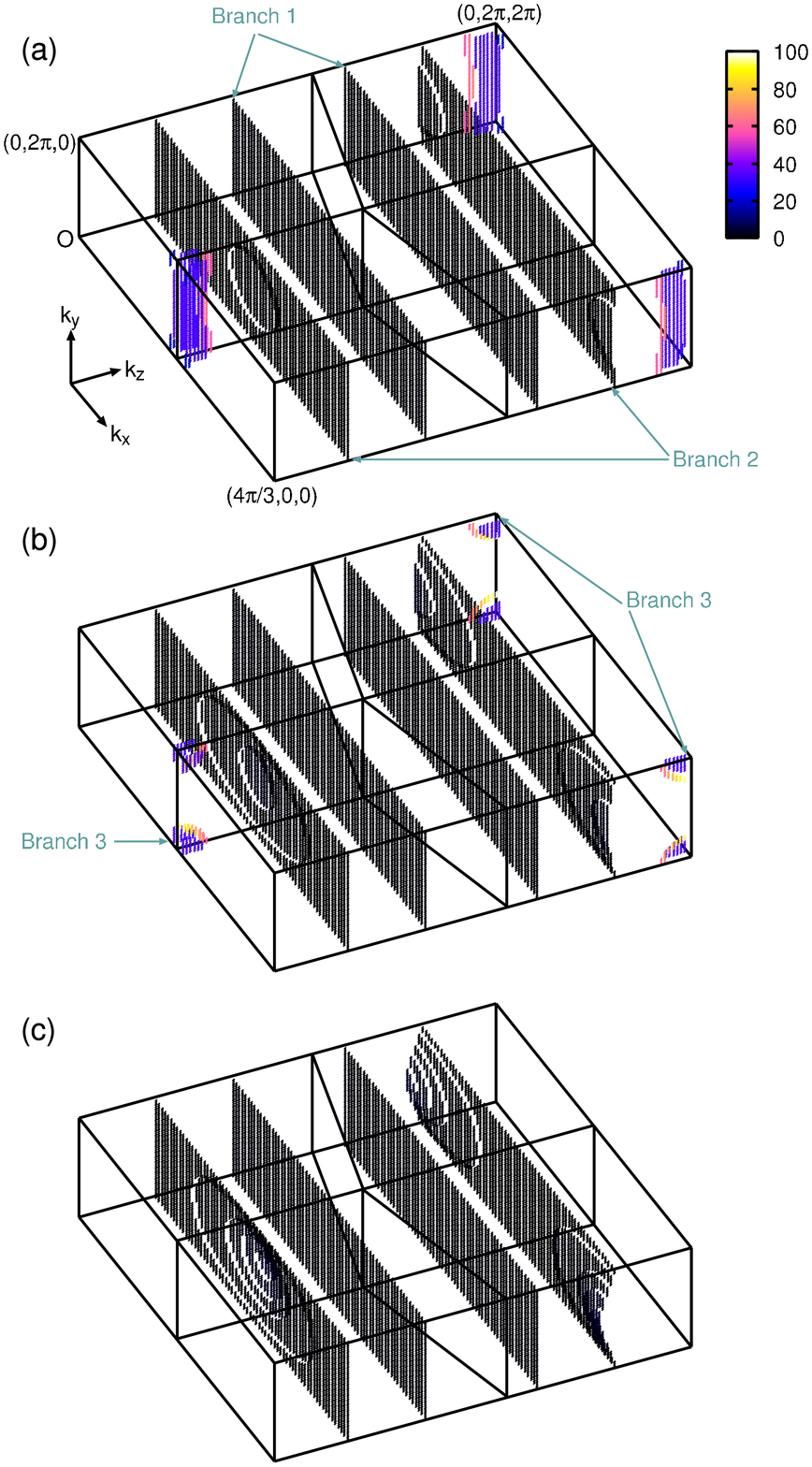}
\caption{\label{figure:6}(Color online) DOS's on Fermi surface for $\Delta V=-0.09\,$eV: (a) for $t_\perp=0.010\,$eV, 
(b) for $t_\perp=0.030\,$eV, and (c) for $t_\perp=0.050\,$eV.}
\end{figure}

On the other hand, for $\Delta V=-0.10\,$eV, $\log_e \Delta_{\mathrm{sc}}$ increases with an increase in $t_\perp$.  
For $\Delta V=-0.09\,$eV, $\log_e \Delta_{\mathrm{sc}}$ decreases with an increase in $t_\perp$, 
but it tends to be much larger than those in the other cases. 
The fact that the behavior of $\log_e \Delta_{\mathrm{sc}}$ differs depending on $\Delta V$ can be explained 
by the configuration of the Fermi surface and the density of state (DOS) on it. Let us show the DOS's on the Fermi surface 
for $\Delta V=-0.12,-0.11,-0.10,$ and $-0.09\,$eV 
in Figs.~\ref{figure:3}--\ref{figure:6}, respectively.
For $\Delta V=-0.12,-0.11,$ and $-0.10\,$eV, we have two quasi-one-dimensional (Q1D) branches of the Fermi surface, 
which we call ``Branch 1'' and ``Branch 2'', as shown in Figs.~\ref{figure:3}--\ref{figure:5}, respectively. 
Branch 2 becomes increasingly warped as $t_\perp$ increases, 
and simultaneously the DOS on the branch is enhanced. A large DOS is favorable for superconductivity. 
However, the warped Fermi surface makes Fermi surface nesting worse. 
The effective pair scattering $V_{ij}(\mib{q})$, defined by 
eqs.~(\ref{eq:1}) and (\ref{eq:7}), is mainly enhanced by Fermi surface nesting; thus, 
the warped Fermi surface suppresses pair instability. Owing to these two conflicting effects 
on the superconductivity, the $\log_e \Delta_{\mathrm{sc}}$ values for $\Delta V=-0.12$ and $-0.11\,$eV 
have maxima at approximately $t_\perp \approx 0.035$ and $0.030\,$eV, respectively, when $t_\perp$ varies. 
Although this situation on the Fermi surface is common for $\Delta V=-0.10\,$eV, 
the DOS on Branch 2 is enhanced more rapidly with $t_\perp$, as shown in Fig.~\ref{figure:5} 
(note that the color bar scale in Fig.~\ref{figure:5} is about 
twice as large as those in Figs.~\ref{figure:3} and \ref{figure:4}), 
and such a large DOS surpasses the other effect due to the warped Fermi surface in this case. 
This is the reason why $\log_e \Delta_{\mathrm{sc}}$ for $\Delta V=-0.10\,$eV increases with an increase in 
$t_\perp$. On the other hand, for $\Delta V=-0.09\,$eV, 
we have three branches of the Fermi surface. Of these, two are Q1D branches and are called similarly to those 
for $\Delta V=-0.12\,$eV. The other branch of the Fermi surface is a quasi-two-dimensional 
(Q2D) branch, newly labeled ``Branch 3'' in Figs.~\ref{figure:6}(a) and \ref{figure:6}(b). 
Noting that the color bar scale in Fig.~\ref{figure:6} is about $100/3$ times as large as that in Fig.~\ref{figure:5}, 
Branch 3 has an extremely large DOS compared with the other branches regardless of $t_\perp$. 
This is why $\log_e \Delta_{\mathrm{sc}}$ for $\Delta V=-0.09\,$eV always becomes large. However, 
as shown in Figs.~\ref{figure:6}(b) and \ref{figure:6}(c), Branch 3 becomes smaller and finally disappears as $t_\perp$ increases. 
This leads to a decrease in total DOS and the degradation of superconductivity.

When we compare the Fermi surface obtained by our present calculation with that observed by ARPES,\cite{TYoshida2009} 
we find that the results for $\Delta V=-0.12$, $-0.11$, and $-0.10\,$eV are similar. 
In other words, in real Sr14-24-41, the charge imbalance between the ladder and chain is 
too large to have three branches of the Fermi surface, as observed in our results for $\Delta V=-0.09\,$eV. 
Thus, the superconductivity in spin-chain ladder cuprate may be enhanced 
if the large charge imbalance between the ladder and the chain can be resolved, for example, 
by varying the elements of the spacer layer without changing their valences.

\begin{figure}
\includegraphics[width=8.0cm]{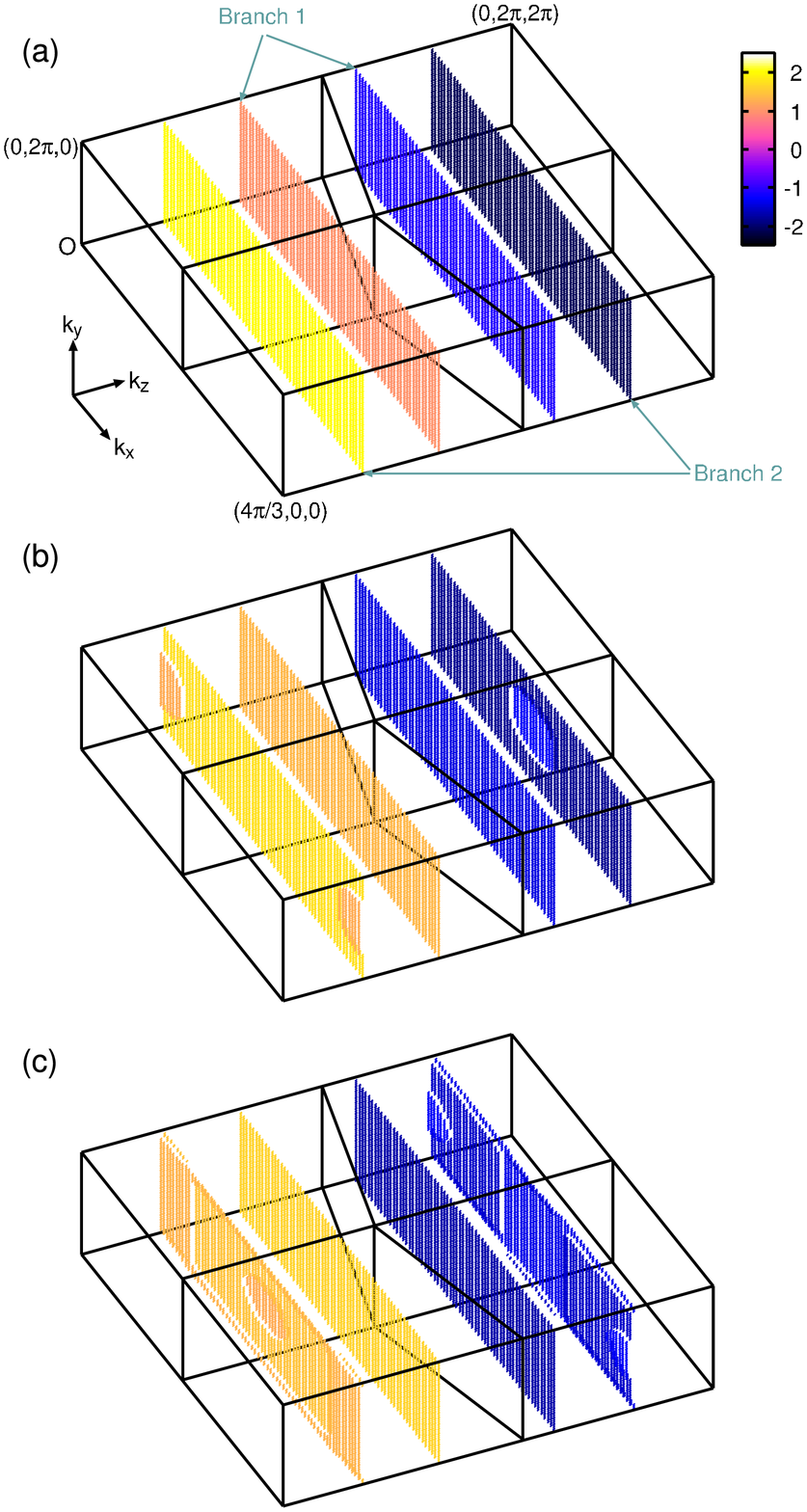}
\caption{\label{figure:7}(Color online) $\Psi_{\mib{k}}^\lambda$ on the Fermi surface for $\Delta V=-0.12\,$eV: (a) for $t_\perp=0.010\,$eV, 
(b) for $t_\perp=0.030\,$eV, and (c) for $t_\perp=0.050\,$eV.}
\end{figure}

\begin{figure}
\includegraphics[width=8.0cm]{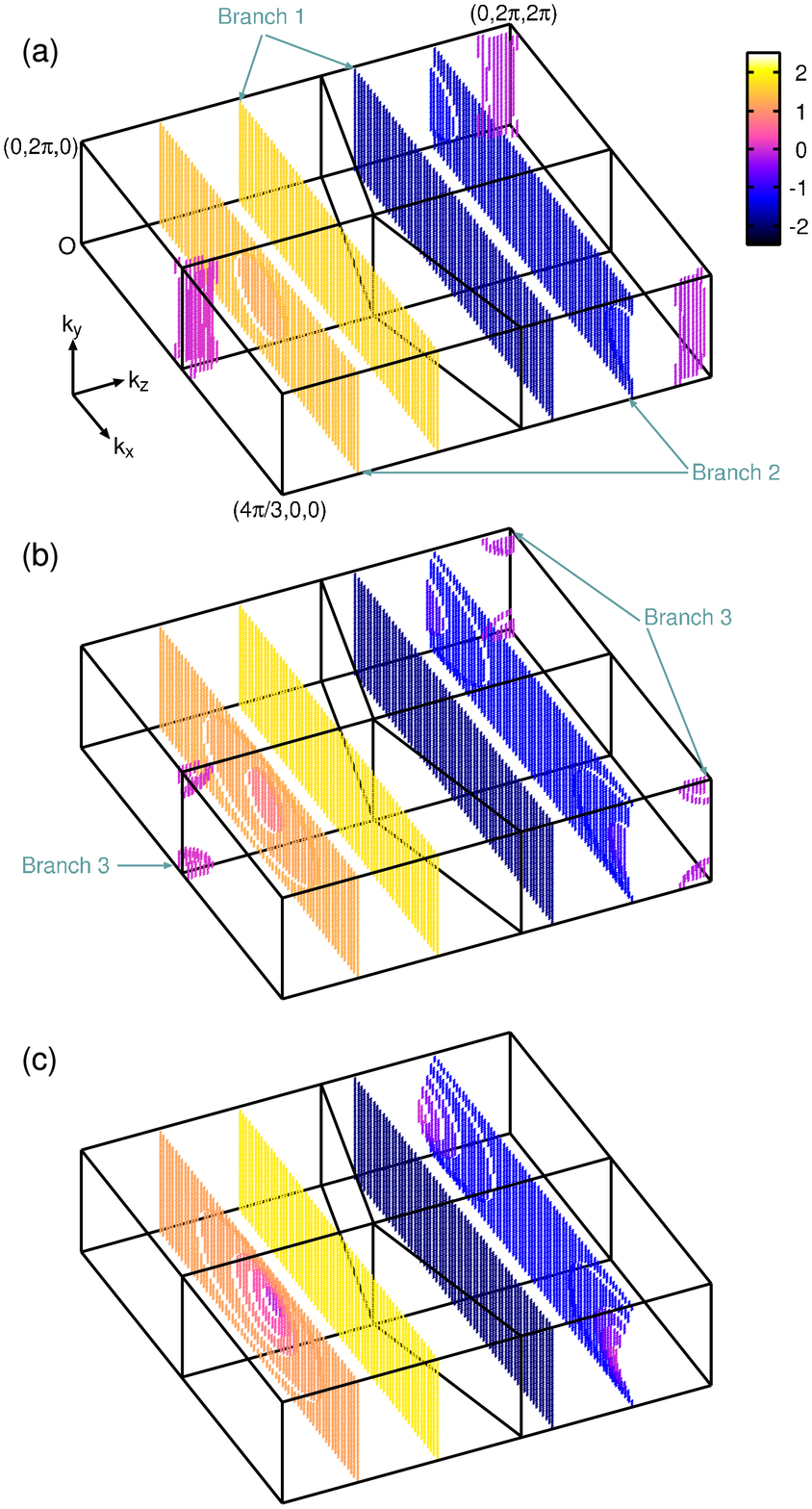}
\caption{\label{figure:8}(Color online) $\Psi_{\mib{k}}^\lambda$ on the Fermi surface for $\Delta V=-0.09\,$eV: (a) for $t_\perp=0.010\,$eV, 
(b) for $t_\perp=0.030\,$eV, and (c) for $t_\perp=0.050\,$eV.}
\end{figure}

Hereafter, we discuss the gap functions of the $p_z$-wave-like superconducting state in detail. 
$\Psi_{\mib{k}}^\lambda$ on the Fermi surface for $\Delta V=-0.12\,$eV and for $\Delta V=-0.09\,$eV 
are shown in Figs.~\ref{figure:7} and \ref{figure:8}, respectively.
For $\Delta V=-0.12\,$eV, $\Psi_{\mib{k}}^\lambda$ changes its sign on the disconnected parts of Branch 1, 
and it does so on those of Branch 2, 
as shown in Fig.~\ref{figure:7}. This is similar to the triplet state derived by Sasaki et al.\cite{Sasaki2004}
Furthermore, $\Psi_{\mib{k}}^\lambda$ has no nodes on the connected parts of each branch. 
Thus, this superconducting state is expected to behave as a fully gapped state for the microscopically experimental probe. 
Actually, the relaxation rate $T_1^{-1}$ of the NMR measurement suggests that the superconducting 
state of Sr14-24-41 at pressures of $3.5$-$3.8$GPa has an $s$-wave-like character.\cite{Fujiwara2003,Fujiwara2009} 
Moreover, in their works, the Knight shift of $^{63}$Cu nuclei for the ladder derived from high fields 
shows no change below $T_{\mathrm{c}}$. This fact strongly suggests that the superconducting state in real Sr14-24-41 can 
be a singlet-Fulde-Ferrell-Larkin-Ovchinnikov (FFLO) state or a triplet superconducting state, 
and the latter is consistent with our result for $\Delta V=-0.12\,$eV. 

However, the fully gapped superconductivity for $\Delta V=-0.12\,$eV does not remain for $\Delta V=-0.09\,$eV. 
As shown in Fig.~\ref{figure:8}, $\Psi_{\mib{k}}^\lambda$ has nodes on Branch 3 or 2. 
These results indicate that our $p_z$-wave-like gap function is not robust in terms of the absence of nodes. 
If the NMR measurement can proceed in Sr14-24-41 under different conditions, we may 
observe $T_1^{-1}$ for the gapless superconductivity. 

Finally, note that the amplitude of $\Psi_{\mib{k}}^\lambda$ on Branch 3 is 
smaller than that on Branch 1 or 2, as shown in Fig.~\ref{figure:8}. 
The large DOS on Branch 3 mainly enhances the superconductivity on the other branches, not on itself. 
This synergistic effect is caused by the interband interaction, 
i.e., $V_{ij}(\mib{k}+\mib{k}^\prime)z_i^\lambda(\mib{k})z_j^\nu(\mib{k}^\prime)$ for $\lambda \neq \nu$ in eq.~(\ref{eq:3}), 
originating from the mixing between d- and p-orbitals. Thus, the hybridization effect also 
plays a significant role in enhancing the superconductivity in this material.

\section{Conclusions}
We have demonstrated that the 3D d-p model with the Q1D structure similar to Sr14-24-41 can represent 
$p_z$-wave-like triplet superconductivity up to the second order in the perturbation theory framework. 
On the basis of this model, we can reproduce the $T_{\mathrm{c}}$ dependence on pressure by changing the interlayer 
coupling. The calculated results on the Fermi surface configuration and the superconducting state can give a 
comprehensive picture to explain the ARPES and NMR experimental results. 
Moreover, our results show the possibility of enhancing the superconductivity 
if the charge imbalance between the ladder and the chain can be decreased by varying the 
elements of the spacer layer. This speculation is based on the fact that 
the hybridization effect due to interlayer coupling 
and the other mixing between d- and p-orbitals is crucial to enhance the superconductivity.

\section*{Acknowledgments}
The authors are grateful to Professors K. Yamaji and I. Hase for helpful discussions. 
The authors also thank the referee for invaluable comments.

\end{document}